\documentclass[aps,prl,twocolumn,amsmath,amssymb,superscriptaddress,nobibnotes,showpacs,floatfix]{revtex4-1}
\usepackage{graphicx}

\usepackage{amsmath}
\usepackage{amssymb}
\usepackage{amsthm}

\usepackage{tabularx}
\usepackage{multirow}

\usepackage{hyperref}
\hypersetup{
  colorlinks = true, 	
  allcolors = blue
}

\newcommand{\dirimg}{./}

\begin{document}

\title{Interplay between adsorbates and polarons: CO on rutile TiO$_2$(110)}

\author{Michele Reticcioli}
\affiliation{University of Vienna, Faculty of Physics and Center for Computational Materials Science, Vienna, Austria}
\author{Igor Sokolovi\'c}
\affiliation{Institute of Applied Physics, Technische Universit\"at Wien, Vienna, Austria}
\author{Michael Schmid} 
\affiliation{Institute of Applied Physics, Technische Universit\"at Wien, Vienna, Austria}
\author{Ulrike Diebold}
\affiliation{Institute of Applied Physics, Technische Universit\"at Wien, Vienna, Austria}
\author{Martin Setvin}
\email{setvin@iap.tuwien.ac.at}
\affiliation{Institute of Applied Physics, Technische Universit\"at Wien, Vienna, Austria}
\author{Cesare Franchini}
\email{cesare.franchini@univie.ac.at}
\affiliation{University of Vienna, Faculty of Physics and Center for Computational Materials Science, Vienna, Austria}

\date[Dated: ]{\today}

\begin{abstract} 
Polaron formation plays a major role in determining the structural, electrical and chemical properties of ionic crystals. 
Using a combination of first principles calculations and scanning tunneling microscpoy/atomic force microscopy (STM/AFM), we analyze the interaction of polarons with CO molecules adsorbed on the rutile TiO$_2$(110) surface. 
Adsorbed CO shows attractive coupling with polarons in the surface layer, and repulsive interaction with polarons in the subsurface layer. 
As a result, CO adsorption depends on the reduction state of the sample. 
For slightly reduced surfaces, many adsorption configurations with comparable adsorption energies exist and polarons reside in the subsurface layer. 
At strongly reduced surfaces, two adsorption configurations dominante: either inside an oxygen vacancy, or at surface Ti$_{5c}$ sites, coupled with a surface polaron. 
\end{abstract}

\maketitle

\begin{figure*}[t!]
    \begin{center}
        \includegraphics[width=1.9\columnwidth,clip=true]{\dirimg 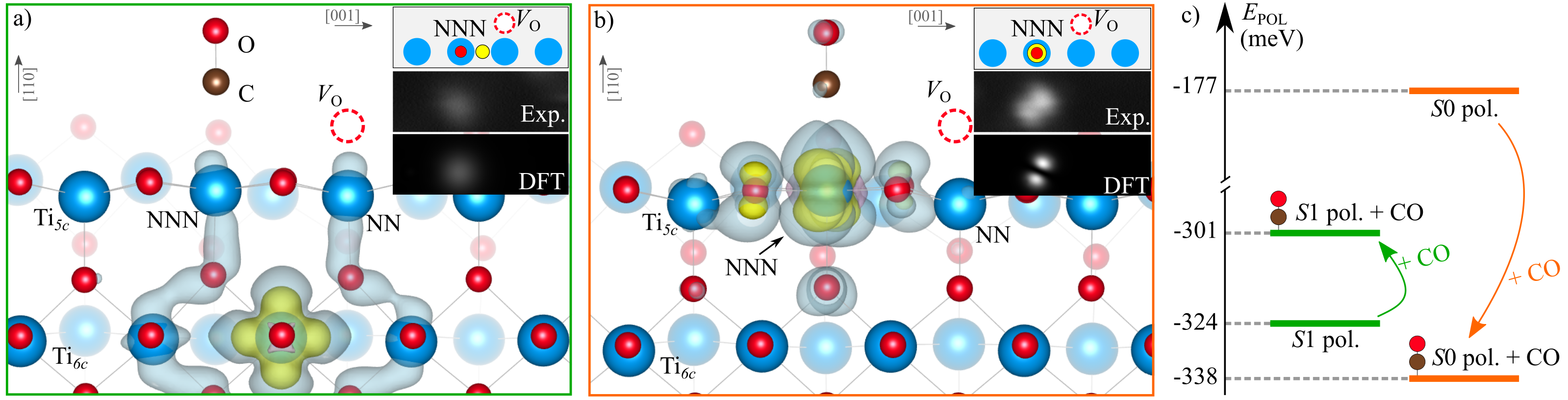}
    \end{center}
\caption{Effects of a CO molecule adsorbed at NNN-Ti$_{5c}$ site on polaronic states at low reduction level (5.6\%, i.e., one ${V_{\rm O}}$ in a 9$\times$2 two-dimensional unit cell). (a,b): electronic charge density of the $S1$ (a) and $S0$ (b) polarons in presence of CO. Atoms at the back are depicted by blurred spheres. A top view of the considered configuration is also sketched in each panel. The insets represent the experimental and simulated STM images.
(c): Polaron formation energy of $S0$ and $S1$ polarons, in case of a TiO$_2$(110) surface with and without adsorbed CO.
}
\label{fig:polCO}
\end{figure*}

A wide range of materials form polaronic in-gap states upon injection of extra charge, as the excess electrons or holes couple to the lattice phonon field. 
The charge carrier generated by defects~\cite{Finazzi2009,Mao2013,Moser2013,Setvin2014,Deak2014,Verdi2017}, doping~\cite{Crevecoeur1970,Janotti2013a,Hao2015a}, adsorbates~\cite{DiValentin2006,Deskins2009,Sezen2015a}, or irradiation~\cite{Sezen2015,Freytag2016,Carneiro2017,Miyata2017}, interact with the lattice field to different extents depending on the electron-phonon coupling, which is strongly material-dependent~\cite{JTDbook,cfGiustino2017,Setvin2014}.
The formation of polarons prevents doping-driven insulator-to-metal transition and dramatically alters the properties of the system and its functionalities~\cite{cfMott2001, JTDbook, Stoneham2007}.
In the strong short-range coupling limit, localized (so called small) polarons form; they locally distort the lattice and lead to the formation of sharp in-gap states~\cite{Stoneham1989,Shluger1993}.
At low temperature, the ground state is determined by the polaronic configuration that minimizes the energy of the system~\cite{Deskins2011,ReticcioliMD}, but even small thermal energies can activate polaron hopping to different hosting sites, thereby changing the nature and properties of the polaronic state~\cite{Dohnalek2010,Kowalski2010,Hao2015a,Reticcioli2017c}.

Formation of polarons is particularly favorable in transition-metal oxides and is further promoted in the vicinity of the surface, where the crystal lattice is more flexible~\cite{Kruger2008,Yoon2015b,Shibuya2017b,Reticcioli2017c}.
Here, we address the interactions between electron polarons and adsorbates.
We show that adsorbates are able to alter the stability of polarons and, in turn, the polarons affect the energetics and configuration of the adsorbates.
We considered CO molecules adsorbed on rutile titanium dioxide, an archetypal polaronic material~\cite{Bogomolov1968}.
In clean (i.e. without CO molecules) TiO$_2$(110) the creation of oxygen vacancies ($V_{\rm O}$) promotes the formation of polarons at the sixfold-coordinated Ti$_{6c}$ atoms lying in the subsurface layer ($S1$)~\cite{Diebold2010a,Deskins2011,Henderson2011,Shibuya2014,ReticcioliMD}. Due to the  attractive interaction, polarons tend to populate sites near $V_{\rm O}$~\cite{Shibuya2014,Yim2016a,Moses2016}.
Polaron hopping from $S1$ to the surface layer ($S0$) is highly unfavorable but might eventually occur at elevated temperatures~\cite{Janotti2013a, Kowalski2010, Setvin2014, ReticcioliMD}.

The effect of polarons is usually not considered in adsorption studies.
CO adsorption on the rutile (110) surface is a well-studied phenomenon from both theoretical and experimental points of view \cite{Diebold2003,PratesRamalho2017a,Mu2017,Zhao2009b,Xu2012a,Cao2017a,Petrik2012a}, yet controversies appear even in elementary issues.
Beyond a general consensus on the local geometric properties (CO molecules adsorb vertically at  Ti$_{5c}$ sites at low coverage)~\cite{Diebold2003,PratesRamalho2017a}, conflicting outcomes have been reported, which either suggest~\cite{Mu2017} or exclude~\cite{Zhao2009b,Xu2012a,Cao2017a} the possibility of CO adsorption at $V_{\rm O}$ sites, in the latter case endorsing a major role played by the fivefold-coordinated Ti$_{5c}$ sites at $S0$.
We show that the apparent disagreements in the literature can be resolved by a proper treatment of polarons and their interaction with adsorbates, by combining density functional theory (DFT) simulations~\cite{Kresse1996a,Kresse1996} with scanning tunneling microscopy (STM) and atomic force microscopy (AFM).

We start by showing how the presence of CO alters the stability and orbital topology of polarons.
We consider one CO molecule adsorbed at the Ti$_{5c}$ site next nearest neighbor to the $V_{\rm O}$ (NNN-Ti$_{5c}$) for a $V_{\rm O}$ concentration of 5.6\% (see Methods~\cite{supplmat} for details). 
The spatial extension of the $S1$ and the $S0$ polaron electronic charge is shown in Fig.~\ref{fig:polCO}(a) and~\ref{fig:polCO}(b), respectively.
The $S1$ polaron retains the same characteristic spatial distribution as in the absence of adsorbates~\cite{Shibuya2012,ReticcioliMD}, with a $d_{z^2}$-$d_{x^2-y^2}$-like orbital character at the hosting Ti$_{6c}$ site and about 1/3 of the charge distributed on the surrounding atoms.
The CO does not alter this situation, and very little (0.1\%) polaronic charge is transfered to the CO molecule. This can be observed in filled-state STM images [see the weak circular spots in the inset of Fig.~\ref{fig:polCO}(a)].

Conversely, a polaron in the $S0$ layer is strongly affected by the CO molecule [Fig.~\ref{fig:polCO}(b)].
The $S0$ polaron acquires a stronger $d_{yz}$ ($54\%$) character as compared to the case with no adsorbates ($43\%$)~\cite{ReticcioliMD}, and a non negligible portion of the polaronic charge ($1\%$) is transfered to the $2\pi$* antibonding orbital of the adsorbed CO~\cite{Bagus1983, Hadjiivanov2002}.
This causes the formation of a double-lobed polaronic cloud above the CO molecule [see the STM images in the inset of Fig.~\ref{fig:polCO}(b)], which has a distinctly different shape from the one shown in  Fig.~\ref{fig:polCO}(a).
Remarkably, the CO-polaron interaction affects strongly the polaron formation energies ($E_{\rm POL}$), as shown in Fig.~\ref{fig:polCO}(c).
While in the $S1$ case the polaron formation energy is only marginally destabilized by the CO ($\Delta E_{\rm POL}^{S1}=+23$~meV), the formation of an $S0$ polaron becomes much more favorable ($\Delta E_{\rm POL}^{S0}=-161$~meV):
The adsorbate changes the polaronic ground state of the system, an effect that was overlooked in previous studies~\cite{FarnesiCamellone2011a,Yu2015a}.
By inspecting the contributions to the polaron formation energy (determined by the balance between the electronic energy gain due to the electron-phonon coupling and the energy cost to locally distort the lattice~\cite{Janotti2013a}) we find that the presence of the CO reduces significantly the structural energy cost in the $S0$ polaron case.
Moreover, we verified that the formation of an $S0$ polaron at the CO-adsorption NNN-Ti$_{5c}$ site is more favorable than any other configuration, by considering different adsorption and polaronic sites.
However, although the CO adsorption reduces the energy barrier for the $S1$-to-$S0$ polaron hopping by 70~meV, the $S1$-to-$S1$ hopping remains predominant (see Fig.~S1~\cite{supplmat}).

\begin{figure}[t]
    \begin{center}
        \includegraphics[width=1\columnwidth,clip=true]{\dirimg 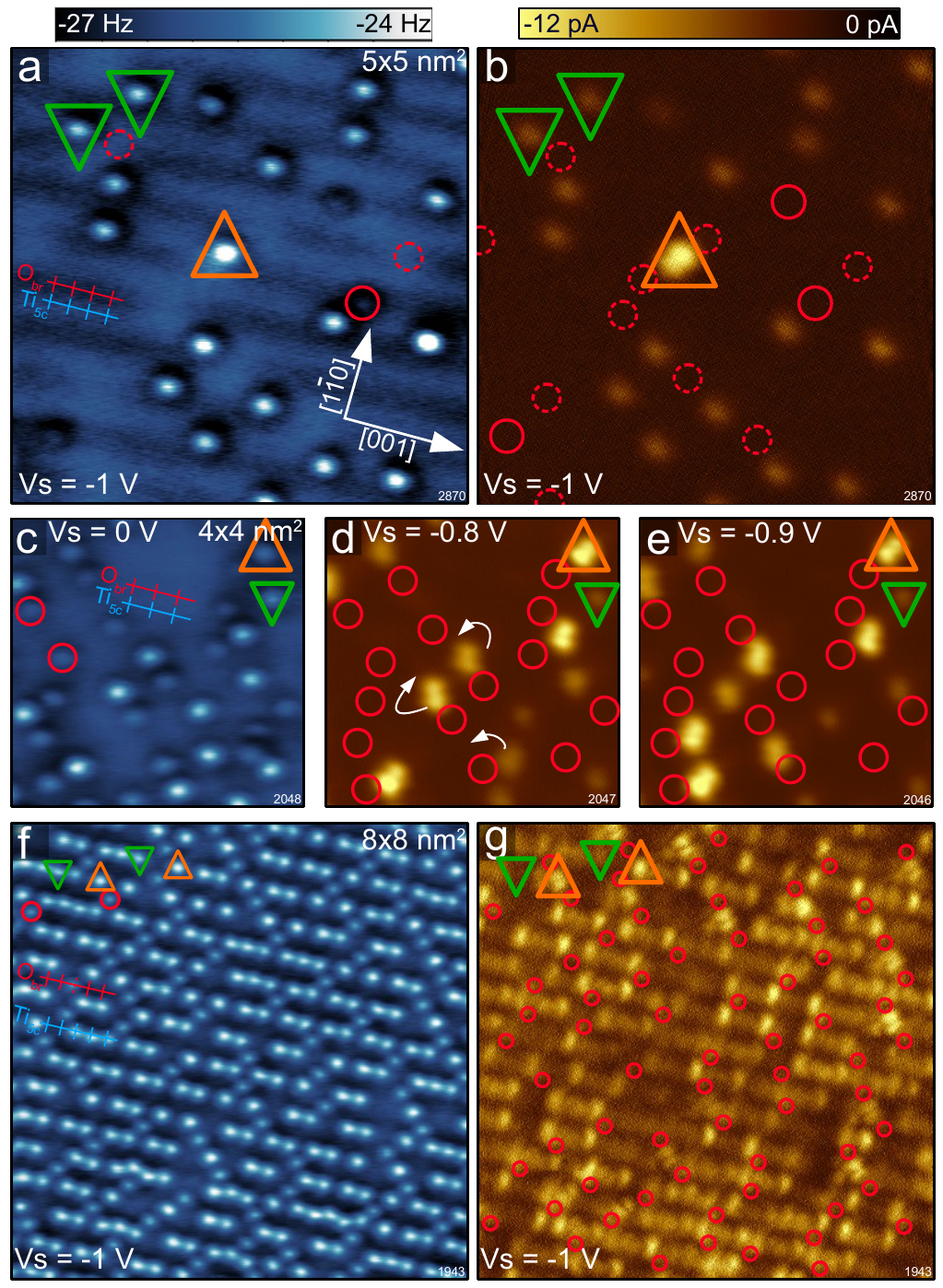}
    \end{center}
\caption{Experimental AFM (blue-white color gradient) and filled state STM images (yellow-black color) of CO adsorbed on the rutile (110) surfaces at different surface reduction states.
Dashed circles show the positions of $V_{\rm O}$s, solid circles show $V_{\rm O}$s with an adsorbed CO molecule, triangles show CO molecules adsorbed on  Ti$_{5c}$ atoms. Down-pointing triangles indicate coupling with $S1$ polarons, up-pointing triangles indicate coupling with $S0$ polarons.
(a,b): Low CO coverage and low reduction level ($\theta_{\rm CO}=0.09$~ML and $c_{V_{\rm O}}=5.8\%$).
(c,d,e): moderate CO coverage and high reduction level ($\theta_{\rm CO}=0.15$~ML and $c_{V_{\rm O}}=14.5\%$). The STM images (d,e) were measured sequentially, and the arrows indicate the diffusion of CO along Ti$_{5c}$ sites, accompanied by polaron hopping.
(f,g): high CO coverage and high reduction level ($\theta_{\rm CO}=0.7$~ML and $c_{V_{\rm O}}=14.5\%$).
}
\label{fig:afmstm}
\end{figure}


Next we focus on the evolution of the CO adsorption process as a function of the $V_{\rm O}$ concentration ($c_{V_{\rm O}}$) and CO coverage ($\theta_{\rm CO}$), see Figure~\ref{fig:afmstm}. We use AFM imaging with a CO-terminated tip to locate the adsorbed CO molecules.
Compared to more commonly used empty-state STM imaging, AFM disturbs the system less (tip-induced CO hopping can be avoided) and provides better resolution of the adsorption geometry. 
Complementary to the AFM, filled-state STM imaging provides information about the polarons.  

Figure~\ref{fig:afmstm}(a,b) shows AFM and STM images of a slightly reduced ($c_{V_{\rm O}}=5.8\%$) surface with a low CO coverage ($\theta_{\rm CO}$=0.09 ML). The CO molecules are predominantly adsorbed on Ti$_{5c}$ sites (brighter spots in the AFM image), and less frequently at $V_{\rm O}$ sites (marked as CO+${V_{\rm O}}$).
The filled-state STM image of the same region [Fig.~\ref{fig:afmstm}(b)] shows mostly weak circular spots at the CO molecules on Ti$_{5c}$ sites, which we attribute to the electronic cloud of $S1$ polarons in the vicinity of the CO, similar to those predicted in the inset of Fig.~\ref{fig:polCO}(a).
There is only one intense double-lobed feature in the STM image, sandwiched by two oxygen vacancies. We attribute this to a CO at the NNN-Ti$_{5c}$ site, coupled to an $S0$ polaron (CO-$S0$-polaron complex), as predicted in Fig.~\ref{fig:polCO}(b).
We note that the CO adsorbed at $V_{\rm O}$ sites observed by AFM carries no in-gap state, and is transparent to the filled-state STM imaging.

At a more reduced surface ($c_{V_{\rm O}}=14.5\%$ and $\theta_{\rm CO}$= 0.15~ML), the AFM image shows that all $V_{\rm O}$ sites are occupied by CO molecules [Fig.~\ref{fig:afmstm}(c)].
The filled-states STM images of the same region show many double-lobed CO molecules at Ti$_{5c}$ sites, indicating the formation of CO+$S0$-polaron complexes [Figs.~\ref{fig:afmstm}(d,e)].
Figs.~\ref{fig:afmstm}(d,e) show two consecutive STM images of the same area, showing diffusion of several CO molecules, associated with the change of their coupling to $S0$ and $S1$ polarons (indicated by arrows in the figure).
The increased number of the CO+$S0$-complexes on the strongly reduced surface is in line with the previously reported polaron dynamics~\cite{Reticcioli2017c}:
The $S1$-to-$S0$ polaron hopping is more favorable due to the repulsive polaron-polaron interactions in the $S1$ layer and attraction of polarons to the $V_{\rm O}$ centers~\cite{ReticcioliMD}.

Finally, by further increasing the CO concentration up to $\theta_{\rm CO}=0.7$~ML~\cite{cov} at highly reduced condition ($c_{V_{\rm O}}=14.5\%$), 
the AFM image clearly shows that CO tends to avoid adsorption sites nearest neighbor to $V_{\rm O}$ [Fig.~\ref{fig:afmstm}(f)], and preferentially populates NNN-Ti$_{5c}$ sites in combination with the $S0$ polaron [double-lobed features in Fig.~\ref{fig:afmstm}(f), particularly intense when the CO is adsorbed at a NNN-Ti$_{5c}$ site sandwiched between two $V_{\rm O}$'s].
CO molecules adsorb also at other Ti$_{5c}$ sites, characterized by weaker STM signals, due to the long-range tail of the electronic clouds of CO-$S0$-polaron complexes at neighbor NNN-Ti$_{5c}$ sites.
All oxygen vacancies are occupied by CO molecules, again with no in-gap state found in the STM images.

\begin{figure}[t]
    \begin{center}
        \includegraphics[width=0.9\columnwidth,clip=true]{\dirimg 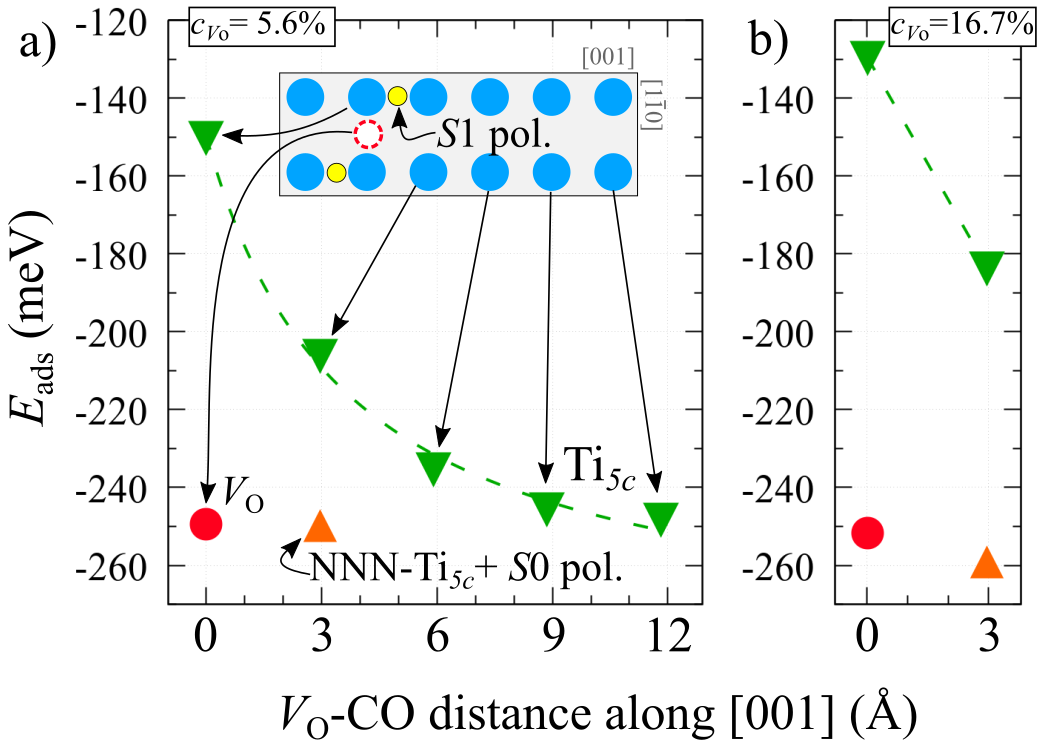}
    \end{center}
\caption{Site dependent adsorption energy.
A CO molecule explores Ti$_{5c}$ sites at various distances from the oxygen vacancy (down pointing triangles), and the $V_{\rm O}$ site (circle), in presence of polarons localized at the $S1$ layer in a reduced slab with $c_{V_{\rm O}}=5.6\%$ (a) and $c_{V_{\rm O}}=17.6\%$ (b). The inset sketches the considered configurations. In addition, we report the case of adsorption at the NNN-Ti$_{5c}$ hosting an $S0$ polaron (up pointing triangle).
}
\label{fig:eads}
\end{figure}

The interpretation of the experimental data is supported by calculated site-dependent CO adsorption energies ($E_{\rm ads}$), see Fig. \ref{fig:eads}.
At low reduction [$c_{V_{\rm O}}=5.6\%$, Fig.~\ref{fig:eads}(a)], the stability of the CO adsorption at non-polaronic Ti$_{5c}$ sites (i.e. CO+$S1$ configurations, down pointing triangles) increases with an increasing distance from the ${V_{\rm O}}$, in accordance with the experiment [see CO+$S1$ circular spots in Fig.~\ref{fig:afmstm}(b)].
CO adsorptions at ${V_{\rm O}}$ (CO+$V_{\rm O}$) or at polaronic NNN-Ti$_{5c}$ (CO+$S0$-polaron complex) sites are essentially degenerate in energy, comparable to the most favorable CO+$S1$-polaron configurations.
In the experiment, the rare occurence of the double-lobed CO+$S0$ spots and CO+${V_{\rm O}}$ features [Figs.~\ref{fig:afmstm}(a,b)] originate from a smaller number of available adsorption sites, as compared to the non-polaronic Ti$_{5c}$ sites~\cite{Mu2017}.

For a strongly reduced surface [$c_{V_{\rm O}}=16.7\%$, Fig.~\ref{fig:eads}(b)], adsorption at Ti$_{5c}$ sites in combination with $S1$ polarons becomes significantly less favorable, mainly due to the absence of Ti$_{5c}$ sites at large distances from the $V_{\rm O}$'s, whereas the CO+$S0$ and CO+$V_{\rm O}$ configurations retain their high stability and represent the most stable solutions.
This is in excellent agreement with the STM measurements that show a progressive increase of double-lobed spots arising from CO+$S0$-polaron complexes with increasing $c_{V_{\rm O}}$ [see Figs.~\ref{fig:afmstm}(c-g)] combined with a large density of CO+$V_{\rm O}$ features [see Figs.~\ref{fig:afmstm}(c,f)].

The interplay between polarons and adsorbed CO molecules significantly affects the adsorption scheme; the different CO-polaron couplings affect the adsorption energies, the bonding distances from the surface as well as the C-O bond length (Fig.~S2~\cite{supplmat}).
The various polaron-CO coupling schemes reported here are consistent with reported experimental data on the CO adsorption: In temperature programmed desorption (TPD), the TPD spectra measured on the rutile (110) surface~\cite{Petrik2012a} are complex, with multiple desorption peaks.
In contrast, the same experiment performed on the anatase TiO$_2$ (101) surface~\cite{Setvin2015} shows a single desorption peak only.
This can be associated with the absence of small polarons at the anatase (101) surface~\cite{Setvin2014}, which simplifies the adsorption scheme in comparison to the polaronic rutile.
Similarly, infrared absorption spectra of CO on the anatase (101) surface always exhibit a single absorption peak~\cite{Setvin2015}, while the rutile (110) surface shows either one or two vibrational states, depending on the reduction level of the crystal~\cite{Xu2012a}.


In summary, by combining first principles calculations and surface sensitive techniques we have elucidated the key role of the interaction between polarons and CO adsorbates on rutile TiO$_2$(110). 
We have shown that CO adsorption promotes polaron transfer from subsurface to surface sites, in particular at highly reduced TiO$_2$ samples, thus reinforcing the activity of surface Ti$_{5c}$ sites.
We have identified three distinct adsorption configurations with well-defined characteristics:
CO at $V_{\rm O}$ sites, CO at Ti$_{5c}$ sites weakly coupled with polarons in the subsurface (manifested by weak circular features in STM), and strongly coupled CO+$S0$-polaron complexes at NNN-Ti$_{5c}$ sites (appearing as double-lobes in STM images).
The coupling between CO and polarons and its interaction with the defective lattice effectively influences the dynamics of the CO+polarons complexes, enabling breaking and recombination of the CO-$S0$/$S1$ complexes.
Our study delivers a consistent and comprehensive picture of the CO adsorption process in an archetypal polaronic material, solves long-standing ambiguities and conflicting interpretations of experimental results, and sets the path for revisiting the interpretation of adsorption processes in polar semiconductors and transition metal oxides.

\section{Acknowledgements}
This work was supported by the Austrian Science Fund (FWF) projects ViCom (F4109-N28) and  POLOX (Grant No. I 2460-N36), by the ERC Advanced Research Grant `OxideSurfaces' and by the FWF Wittgenstein-prize (Z250-N16).
The computational results were achieved by using the Vienna Scientific Cluster (VSC).

\bibliography{bib-2018-06-29}


\end{document}